\documentclass[a4paper,
               nospread,     
               ]{jacow}

\makeatletter%
	\ifboolexpr{bool{xetex}}
	 {\renewcommand{\Gin@extensions}{.pdf,%
	                    .png,.jpg,.bmp,.pict,.tif,.psd,.mac,.sga,.tga,.gif,%
	                    .eps,.ps,%
	                    }}{}
\makeatother

\ifboolexpr{bool{xetex} or bool{luatex}} 
 {}                                      
 {\usepackage[utf8]{inputenc}}           

\usepackage[USenglish]{babel}

\ifboolexpr{bool{jacowbiblatex}}%
 {%
  \addbibresource{jacow-test.bib}
  \addbibresource{biblatex-examples.bib}
 }{}
\title{Using the LH\lowercase{e}C ERL to Generate High-Energy Photons}

\author{N.S.~Mirian\thanks{najmeh.mirian@desy.de}, Deutsches Elektronen-Synchrotron DESY, Germany \\
 E.~Salehi\thanks{elham@ims.ac.jp}, UVSOR, Institute for Molecular Science, Okazaki, Japan\\
 F.~Zimmermann, CERN, Meyrin, Switzerland }

\begin{document}
			
\maketitle

\begin{abstract}
The Large Hadron electron Collider (LHeC) is a proposed future particle physics project colliding 60 GeV electrons from a six-pass recirculating energy-recovery linac (ERL) with 7 TeV protons stored in the LHC. The ERL technology allows for much higher beam current and, therefore, higher luminosity than a traditional linac. The high-current, high-energy electron beam can also be used to drive a free electron laser (FEL). In this contribution, we examine how the LHeC ERL can serve as a source of high-energy photons for studies in nuclear physics, high-energy physics, Axion detection, dark energy, and protein crystallography.  In the first section, we discuss the performance of the LHeC-based FEL, operated in the SASE mode for generating photon pulses at wavelengths ranging from 200 keV to 600 keV. In the second section, we investigate photon production via Laser Compton scattering (LCS).
\end{abstract}

\section{Introduction}
Photon beams with high energies ($>$10~keV) offer novel possibilities for research and application. An intense high-energy photon beam with multiple outstanding 
characteristics, such as energy tunability, good 
directivity, quasi-monochromaticity, etc., can be employed for numerous 
applications in nuclear physics, high energy physics, and non-destructive material 
analysis.
Potential applications of the high photon energy include protein crystallography, coronary angiography, nanotechnology, along with searches for Dark Photons and Axion-like Particles (ALPs). 

\begin{figure}[htbp]
\begin{center}
\includegraphics[width=0.4\textwidth, height=3.8 cm]{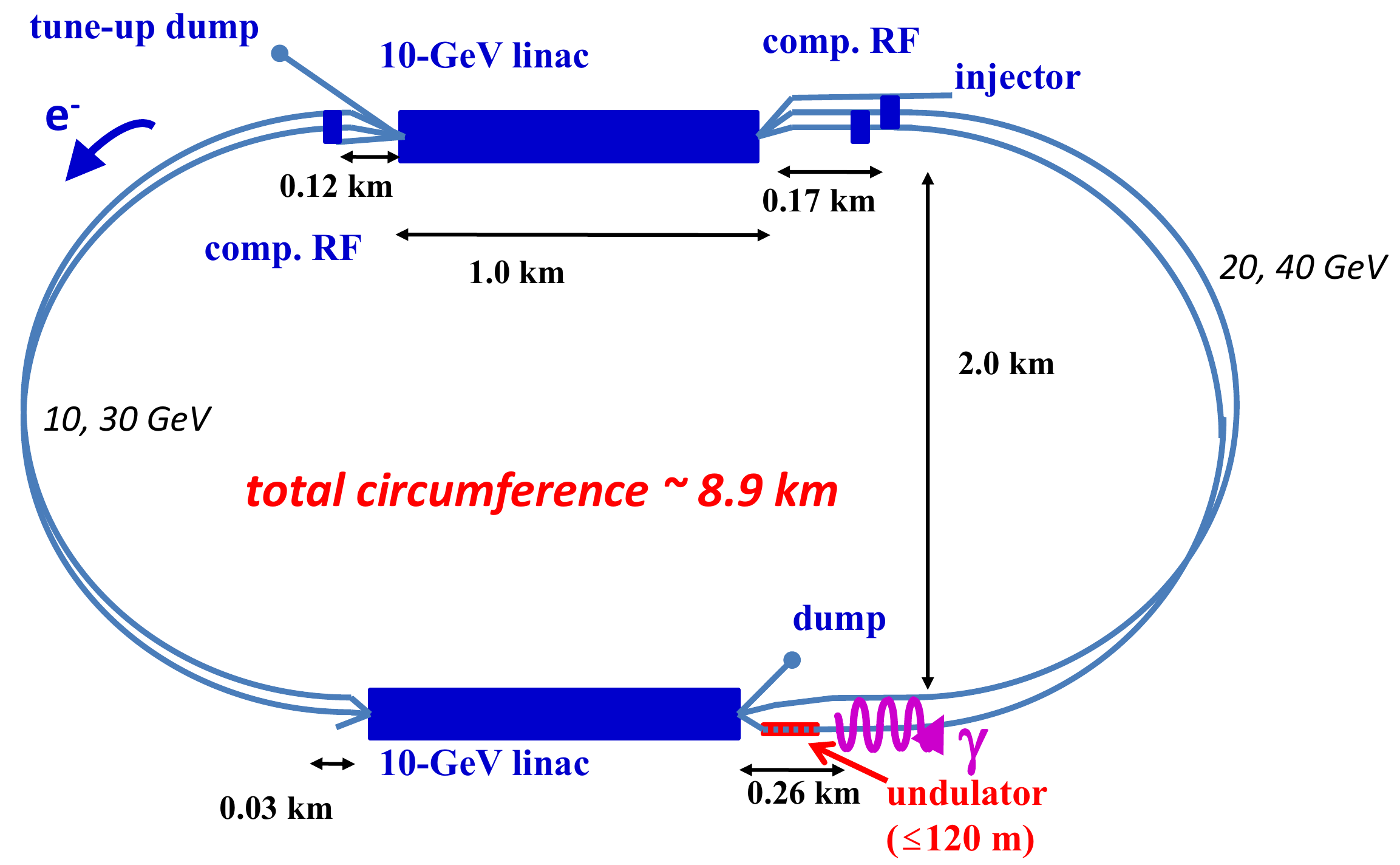}
\end{center}
\caption{\label{fig:scheme} LHeC recirculating linac reconfigured for FEL operation \protect\cite{LHeCFEL}.}
\end{figure}

Recently, in Ref.~\cite{LHeCFEL} we have demonstrated the potential generation and amplification of photons with a few hundred-kilo electron-volt energy by using the high-energy electron beam available at the Large Hadron electron Collider (LHeC). 
The LHeC  is a proposed 
future particle-physics project, colliding 60 GeV electrons from a
six-pass  recirculating energy-recovery linac (ERL) with 7 TeV protons circulating 
in the existing LHC \cite{LHeCdesign,lhechf}. 
The ERL technology allows for much higher beam current and,
therefore, higher luminosity than a traditional linac. 
We considered using the high-current, high-energy electron beam of he LHeC to drive a
free electron laser (FEL).
Figure \ref{fig:scheme} sketches the reconfigured LHeC-FEL. 
We demonstrated that such ERL-based high-energy  
FEL would have the potential to provide orders
of magnitude higher average brilliance at sub \AA\ wavelengths
than any  other FEL either existing or proposed. 
We investigated the performance of an LHeC-based FEL 
operation in the self-amplified spontaneous emission (SASE) mode, using the LHeC electron beam 
after one or two turns of acceleration, with beam energies of, e.g., 10, 20, 30, and 40 GeV, to produce X-ray pulses at wavelengths ranging from 1.5 keV to 24 keV. 
In addition, we explored a possible path to use the 40 GeV electron beam for generating photon pulses at even much higher photon energy up to 100 keV, which would also allow a pioneering step into the picometer wavelength regime \cite{LHeCFEL}.\\
However, photonuclear experiments as well as quantum electrodynamics (QED) and quantum chromodynamics (QCD) experiments for high-energy physics require electromagnetic interactions in at energies extending from several MeV to GeV. 
Such gamma rays can be generated by Laser Compton scattering (LCS) of an intense laser beam off a high-energy electron beam. The LCS technique generates intense gamma-rays exhibiting various remarkable properties, e.g., quasi-monochromaticity, energy tunability, good directivity, and high polarization, which can allow many applications in nuclear physics and non-destructive analysis.

In this article, we demonstrate the potential of LHeC-FEL for generating high photon energy via the FEL and LCS processes.

\section{Reconfigured LH\lowercase{e}C-FEL }
To generate a high photon energy beam, we can utilize the 40 and 30 GeV electron beams of the LHeC.  
As indicated in Fig.~\ref{fig:scheme}, the beam energy of 40 GeV can be attained after two passes through 
the two 10 GeV linacs, instead of the three passes reaching 60 GeV, in the
standard LHeC  mode of operation. 
A beam energy of 30 GeV is also readily obtained after either 1.5 turns or after two turns with appropriate linac voltages and phasing. 

In this study, we used the ELEGANT particle tracking simulation of Ref.~\cite{LHeCFEL}.
Table~\ref{lhecparam} summarizes the 40 GeV electron beam parameters optimized 
for LHeC FEL operation. The bunch compression using three linac passages and 
three arcs increases the peak bunch current by more than an order of
magnitude while preserving a reasonable transverse emittance and energy
spread suitable for FEL operation.

\begin{table}[tbp] 
\caption{\label{lhecparam} The main LHeC-ERL electron beam parameters. Peak
current, bunch length, and transverse emittance were obtained from
a tracking simulation with the code ELEGANT \protect\cite{LHeCFEL, elegant}.}
\centering
\begin{tabular}{lcr}
\hline\hline
Parameters  & Unit  & Value\tabularnewline
\hline 
injection energy  & GeV  & 0.5\tabularnewline
final energy  & GeV  & 40/30 \tabularnewline
electrons per bunch  &  & 3$\times10^{9}$\tabularnewline
initial FWHM bunch length  & $\mu$m  & 234\tabularnewline
final FWHM bunch length  & $\mu$m  & 35 \tabularnewline
initial peak beam current  & kA  & 0.6\tabularnewline
final peak beam current  & kA  & 4 \tabularnewline
final hor.~normalized emittance  & $\mu$m  & 0.9 \tabularnewline
final vert.~normalized emittance  & $\mu$m  & 0.4\tabularnewline
bunch spacing  & ns  & 25\tabularnewline
final rms energy spread  & $\%$  & 0.01\tabularnewline
\hline\hline
\end{tabular}
\end{table}

\section{Generation of a few hundred keV photon energy by FEL }
To produce the high-energy SASE photons with the 40 GeV and 30 GeV electron beams from the second linac of the LHeC, consider our FEL simulation with a ``Delta undulator'' featuring an~18 mm period and 5~mm minimum gap. 
The Delta undulator \cite{DeltaU} can shape the FEL 
photon polarization.  
%
\begin{figure}[htbp]
\begin{center}
\includegraphics[width=0.4\textwidth, height=4 cm]{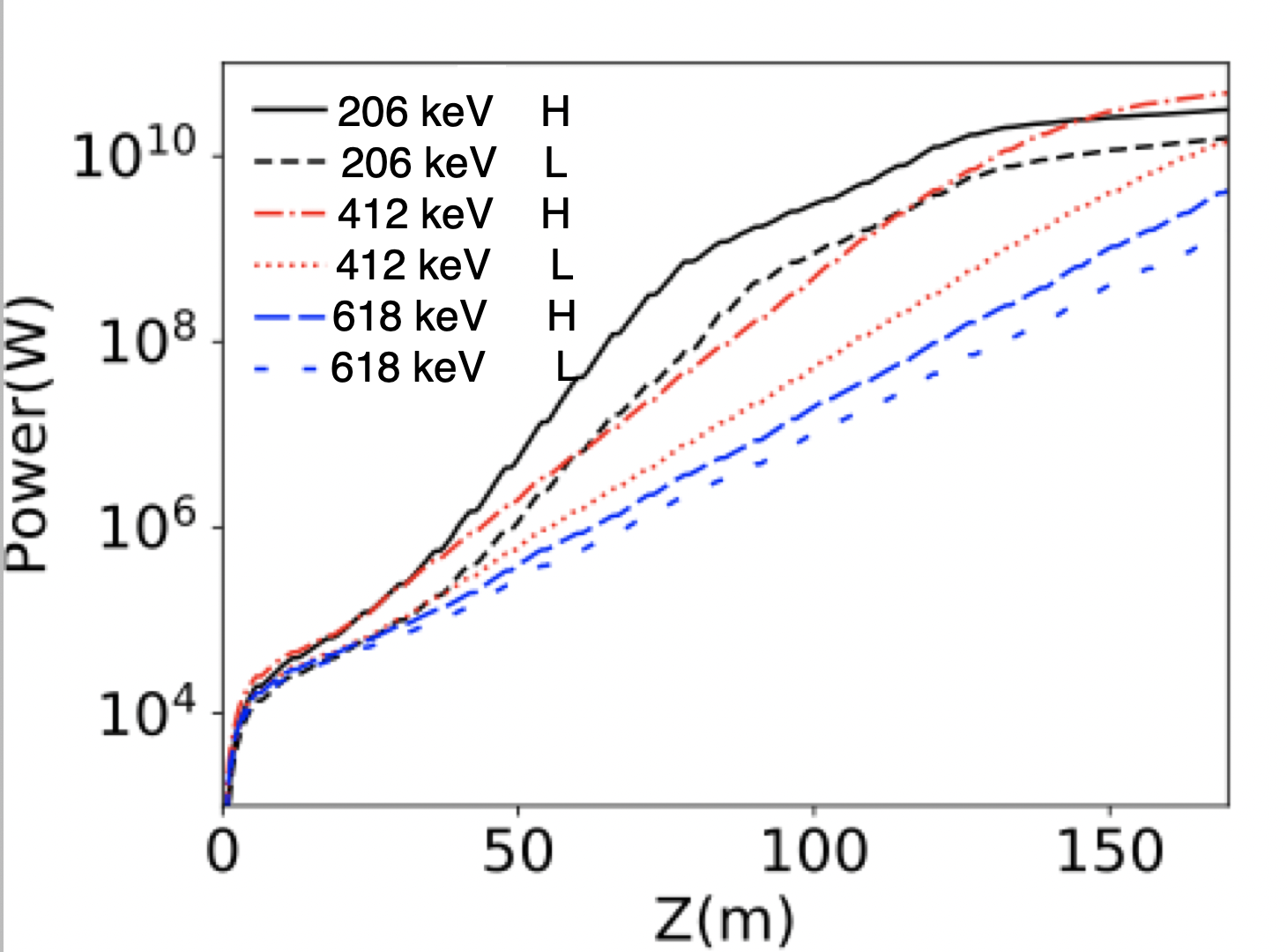}
\end{center}
\caption{ Simulated power growth for cases of helical (H) and linear (L) 
polarization for photon energies higher than 206 keV.  
The simulations studied the SASE amplification for an electron beam of 
either 30 GeV (black lines) or 40 GeV 
(red and blue lines) passing through a helical or planar 
Delta undulator FEL line. The black lines refer to 206 keV, 
the red ones to 412 keV, and the blue lines to 618 keV.
\label{fig:FELpower-growth40Gev}}
\end{figure}

Figure \ref{fig:FELpower-growth40Gev} displays GENESIS code \cite{genesis} simulations for the helical and planar sets up of the Delta undulator FEL line, generating photons with helical and linear polarization, respectively.   
The simulations at a wavelength of 206 keV were performed for a 30 GeV electron beam passing through the Delta undulator,  with a gap of $\sim 6.5$ mm, considering 
either helical or linear polarization,  with results illustrated 
 by the black solid and dashed lines, respectively. 
The figure also presents the radiation amplification at photon energies of 412 and 618 keV, represented by the red and blue lines, respectively, for a 40 GeV beam passing 
through the helical or planar Delta undulator line.

Predictions from the 1D and 3D FEL theories of Refs.~\cite{Saldin2004, Saldin_2010,Xie} are in good agreement with the GENESIS simulation results for wavelengths of a few hundred keV photon energy, presented in Fig.~\ref{fig:FELpower-growth40Gev} \cite{LHeCFEL}.

\section{Generation of a few hundred MeV to few GeV photon energy by LCS }
\begin{figure}[htbp]
\begin{center}
\includegraphics[width=4 cm, height=2.5 cm]{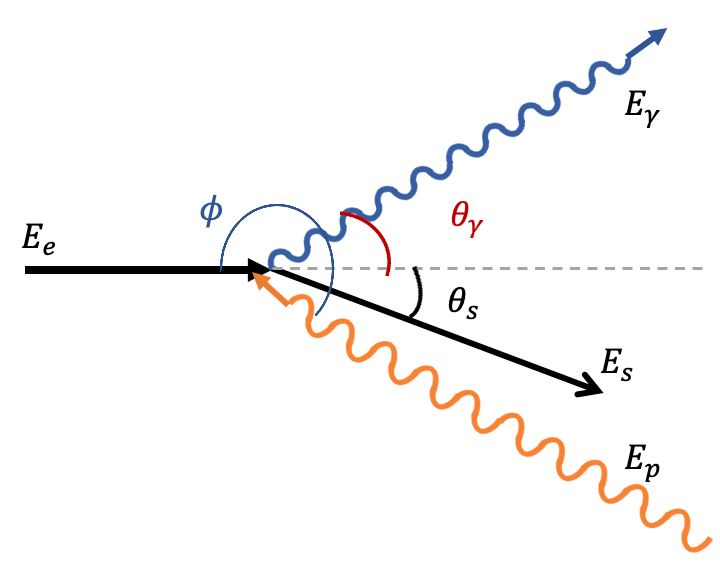}
\end{center}
\caption{Scattering process in the laboratory frame. \label{fig:LCS}}
\end{figure}

The field of high-energy and nuclear physics requires moving into the realm of photon energies even greater than MeV for some specific experiments. 
to this end, Laser-Compton backscattering technique is a unique technique to 
generate photons of energy much higher than a few MeV.   
The schematic of laser Compton scattering is illustrated in Fig.~\ref{fig:LCS}.

A laser photon of energy $E_{p}$ and an electron with kinetic energy $E_{e}$ collide 
at an angle ${\phi}$. After the collision, the photon and the electron are scattered with energy $E_{\gamma}$ and $E_{se}$, respectively. 
Thanks to energy conservation,  the energy of the scattered photon is directly 
related to its scattering angle $\theta_\gamma$, according to 
\begin{equation}
E_{\gamma}=\frac{E_{p}(1-\beta \cos\phi)}{(1-\beta \cos\theta_{\gamma})+\frac{E_{p}}{m_{0}c^{2}}\sqrt{1-\beta^{2}}(1-\cos(\theta_{\gamma}-\phi))},
\end{equation}
where $m_{0}$ denotes  the rest mass of the electron, $\beta$ the ratio of the electron velocity and light velocity $c$, and $\theta_{\gamma}$ the angle of the scattered gamma-ray photon. In the case of a head-on collision, $\phi=\pi$ Eq.~(1) can be simplified to 
\begin{equation}
E_{\gamma}=\frac{(1+\beta)E_{p}}{1-\beta \cos\theta_{\gamma}+(E_{p}/m_{0}c^{2})\sqrt{1-\beta^{2}}(1+\cos\theta_{\gamma})}\; .
\end{equation}

The maximum scattered photon energy, which can be observed in the backward direction of the incident photon, namely at $\theta_{\gamma}=0$, is given by $E_{\gamma}=4\gamma^{2}E_{p}$
with $\gamma$ the Lorentz factor, $\gamma=\sqrt{1-\beta^{2}}$, and $E_p$ the energy of the primary laser photon. As revealed by this equation, the energy of the backward  scattered photons is estimated to be $4\gamma^{2}$ times higher than the energy of the initial laser photons. For LHeC, in the case of a CO$_{2}$ laser with laser photon energy $E_{p}=0.117~$eV, power of $0.5~W$, and beam size $\sigma_{x,y}=5 ~mm$ the maximum energy of the scattered photons is estimated to be a few GeV, 
that is in the high energy of gamma-ray photons.
\begin{figure}[htbp]
\begin{center}
\includegraphics[width=4 cm, height=3. cm]{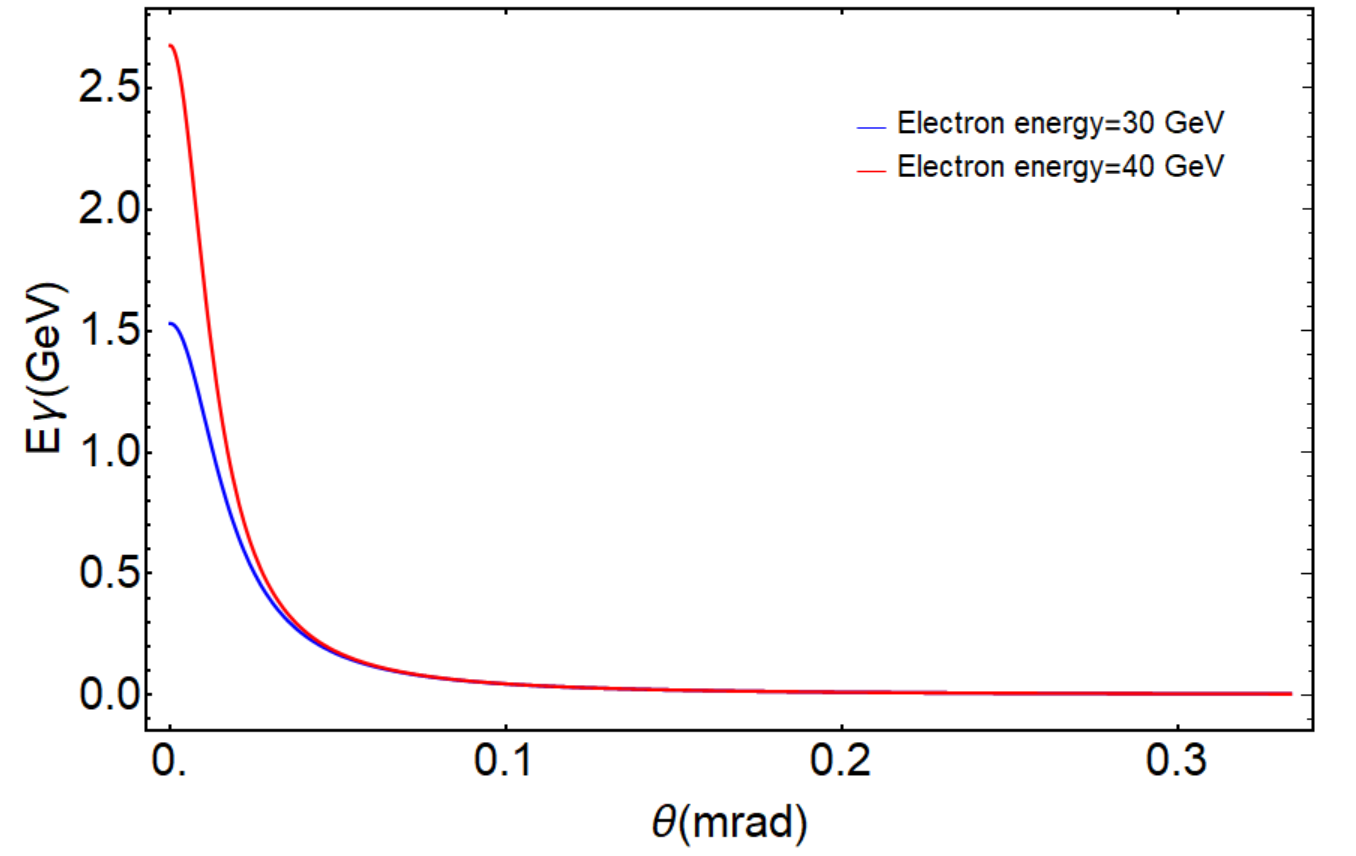}
\includegraphics[width=4 cm, height=3.3 cm]{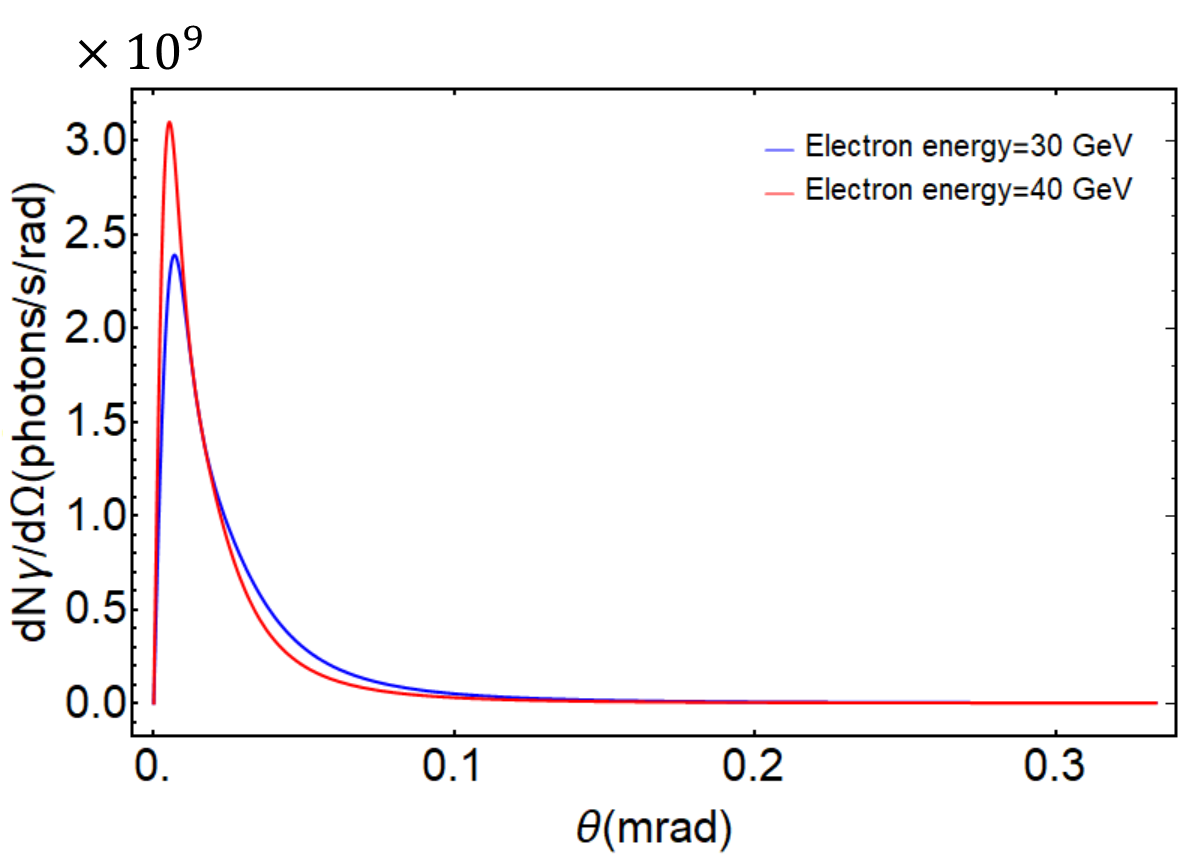}
\end{center}
\caption{Left: Energy distribution of the scattered photons. Right: Rate of gamma-ray photons per second and per radian versus the scattering angle $\theta$,  from a head-on collision of a  10.6 $\mu$m laser beam with an electron energy of 30 GeV (blue) and 40 GeV (red). \label{fig:Egamma}}
\end{figure}

The rate of gamma-ray scattered photons per unit time, $\dot{N}_{\gamma}$ can be evaluated by a numerical calculation of the product of luminosity $L$ and scattering cross-section $\sigma(\theta_{\gamma})$ integrated over  the scattering angle up tp  
$\theta_{\gamma}$ \cite{9}: $\dot{N}_{\gamma}=L\sigma(\theta_{\gamma})$.
The differential cross section for an electron in the laboratory frame can be derived from the Klein–Nishina formula \cite{10} in the electron rest frame followed by a Lorentz transformation. 
In the electron rest frame, the laser photon always collides ``head-on''. Therefore, the differential cross-section is independent  of the incident angle of the laser $\phi$. The differential cross section can be written as \cite{STEPANEK1998174} :
\begin{equation}
\frac{d\sigma}{d\Omega}=\frac{1}{\gamma^{2}(1-\beta\cos\theta)^{2}}\frac{r_{0}^{2}}{2R^{2}}\left( R+\frac{1}{R}-1+\left(\frac{\cos\theta-\beta}{1-\beta\cos\theta}\right)^{2}\right)\; ,
\end{equation}
where $R=\left(1+\frac{\gamma(1+\beta)E_{p}}{m_{0}c^{2}}\left(1+\frac{\cos\theta-\beta}{1-\beta\cos\theta}\right)\right)$ and $r_{0}$ designates the classical electron radius. The luminosity is [9]
\begin{equation}
L=\frac{fN_{e}N_{p}}{2\pi}
\int_{\frac{-L_{\rm int}}{2}}^{\frac{-L_{\rm int}}{2}} \frac{1}{\sqrt{\sigma_{ex}^{2}+\sigma_{Lx}^{2}}\sqrt{\sigma_{ey}^{2}+\sigma_{Ly}^{2}}} \,dz.
\end{equation}
Here, $N_{e}$ is the number of electrons per bunch, $N_{p}$ is the number of laser photons, and $f$ is the collision frequency, while 
$\sigma_{x}$ and $\sigma_{y}$ are the rms horizontal and
perpendicular beam sizes, respectively. 
The subscripts $e$ and $L$ refer to the electron beam and to the laser, respectively. 
From the above equations, the total number of scattered photons per unit of time and per scattered photon energy unit can be written as: 
\begin{equation}
\frac{dN_{\gamma}}{dt\; dE_{\gamma}}=\frac{fN_{e}N_{p}}{2\pi}\frac{L_{\rm int}}{\sqrt{\sigma_{ex}^{2}+\sigma_{Lx}^{2}}\sqrt{\sigma_{ey}^{2}+\sigma_{Ly}^{2}}} \frac{d\sigma}{dE_{\gamma}}\; ,
\end{equation}
where $L_{\rm int}$ is the effective interaction length.

\begin{figure}[htbp]
\begin{center}
\includegraphics[width=0.4\textwidth, height=4 cm]
{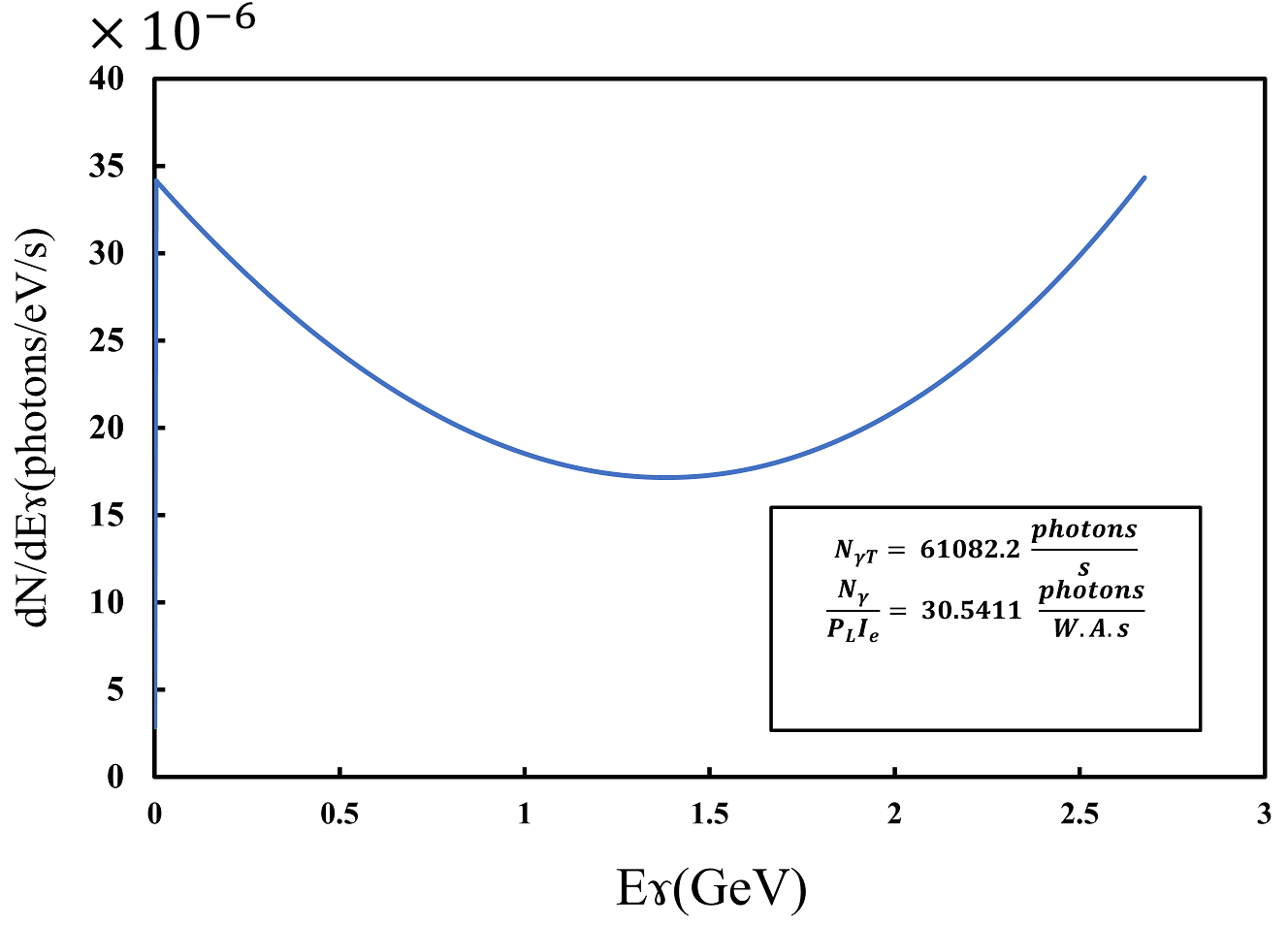}
\end{center}
\caption{ Energy distribution of the  scattered gamma-ray photons versus photon energy produced by the head-on collision of a $\lambda$=10.6~$\mu$m laser beam with a 40 GeV electron beam.}\label{fig:Ngamma}
\end{figure}

Based on Eq.~(2), the energy distribution of the scattered photons can be evaluated as a function of scattering angle,  as is shown in Fig.~\ref{fig:Egamma} (left). Figure \ref{fig:Egamma} shows the energy distribution of scattered gamma-ray photons (left) and the rate of gamma-ray scattered photons per solid angle as a function of the scattering angle (right) for the collision of $0.117$~eV CO$_2$ laser photons with the LHeC electron beam at an energy of 30 GeV (blue) and 40 GeV (red). The figure indicates that most of the high energy scattered photons and most of the photons fall within a very narrow scattering angle. As can also be seen in Fig.~\ref{fig:Egamma}, for higher electron energy 40 GeV, the scattered photon energy is larger.
Using equations (3) and (4), we can calculate the energy spectrum inside an aperture of radius $R=5$~mm at a 
distance $l=8$~m from the center of the collision region extending over 
the length $L_{\rm int}=7$~m. The result is shown in Fig.~\ref{fig:Ngamma}. The spectrum exhibits an edge corresponding to a high-energy cutoff, 
which is determined by the incident electron and photon energies. 
The total gamma-ray flux normalized to laser power and beam current has been evaluated at approximately $30.541$~{\rm photons}/W/mA/s.


\section{Conclusions }
The generation of high-energy photon beams opens the door to novel research and application opportunities. Many applications in nuclear physics, high-energy physics, and non-destructive analysis call for an intense high-energy photon beam with energy tunability, good  directivity, quasi-monochromaticity, and high polarization. 
In this article, we examined the generation of high-quality, high-energy photon beams at the proposed CERN-based LHeC collider, 
via two different approaches, promising the generation of intense x- or gamma-rays with photon energies above 200 keV.  
We assumed the simulated electron beam of Ref.~\cite{LHeCFEL} after two turns 
behind the second linac, with a beam energy of either 30 or 40 GeV. 
Simulation results presented in the first part of this article 
suggest that, by employing the 'Delta undulator' with 18 mm period length, SASE FEL radiation with photon energies in the range of 200-620 keV can be produced. 
In the second part, we described the laser Compton scattering concept, enabling the generation of even much higher photon energies in the range of 400 MeV to 2.5 GeV,
and we discussed the energy-angle correlations in the resulting photon beams.

\end{document}